\documentclass[12pt]{article}

\usepackage{graphics}
\usepackage{amsmath}
\def\ni{\noindent}
\def\ph{{\phantom{...}}}
\def\={\phantom{..} = \phantom{..}}

\def\+{\phantom{..} + \phantom{..}}
\def\>{\phantom{..} > \phantom{..}}
\def\<{\phantom{..} < \phantom{..}}
\def\-{\phantom{..} - \phantom{..}}

\def\vs{{\em vs.}}

\def\bq{\begin{quote}}
\def\eq{\end{quote}}
\def\be{\begin{equation}}
\def\ee{\end{equation}}
\def\bar{\begin{eqnarray}}
\def\ear{\end{eqnarray}}
\def\no{\nonumber}

\def\Sch{Schr{\"o}dinger}
\def\Schism{Schr{\"o}dingerism}

\def\Schists{Schr{\"o}dingerists}

\def\Copists{Copenhagenists}

\def\skN{\sum_{k=1}^N}

\def\hstm{\frac{\hbar^2}{2 m}}
\def\cL{{\cal L}}
\def\dint{\int_0^T\,dt\,\int_A\,\prod\,dx_k}
\def\din{\int_0^T\,dt\,\int_A\,\prod\,dx_k}

\title{\bf Locality, Micro- \vs\ Macro-, Particle Interpretations and All That: \\
a Lagrangian Approach to the Measurement Problem\\[3in]}
\author{W. David Wick\footnote{email: wdavid.wick@gmail.com}}

\begin{document}
\maketitle
\pagebreak

\section*{Abstract}
In 2017, this author proposed, as a resolution of the Measurement Problem, 
that terms be added to \Sch's wavefunction equation, rendering 
it nonlinear.  
Said equation derived from a trick employed by 
S. Weinberg in 1989 which may be unfamiliar to most physicists, as well as uninterpretable
in terms of local (``particle") interactions.
Motivated by A. O. Barut's work on electrodynamics,  
here I analyze which kinds of nonlinear theories can be derived from  
Lagrangian field-theory by integrating out some fields. The issues of ``What is 
a local interaction?" and ``Might there be 
Micro- and Macro-fields?" arise.
In the end, I will argue that my 2017 theory cannot be given a ``particle"
interpretation, nor be derived from a splitting into the two categories of fields.

\section{Introduction\label{introsection}}

For my 2017 proposal, \cite{WickI}, I adopted an elegant trick (utilized by S. Weinberg in his 1989
study of ``Nonlinear Quantum Mechanics", \cite{Weinberg89}) for deriving an evolution equation 
from a functional of the wavefunction representing the system energy.\footnote{``Expected
energy" for \Copists; just ``energy" for \Schists.}
But this approach
may be unfamiliar to most physicists, who are used to field theories being derived
from an action principle and a Lagrangian.

The present work was motivated partially by that concern
and also from a comparison I made, see \cite{Wick_Barut}, of my theory to the work of A. O. Barut
and associates in the 1980s and 1990s on the self-consistent, Dirac-Maxwell electrodynamics.
Of course, the two programs are very different; Barut was interested in an alternative to
the Quantum Electrodynamics of Feynman, Schwinger, Dyson {\em et al}. as an explanation
of phenomena on the atomic level. By contrast, my 2017 attempted to revise \Sch's 1926 theory
at the macro level (of apparatus), without perturbing it by much on the micro level.  
Nevertheless, I noted that, by ``integrating out" some fields, Barut obtained a Lagrangian
superficially similar to that in my theory if given a similar construction.
So I aimed to find a Lagrangian formulation, and then determine whether it might arise
from another---perhaps ``local"?---theory with additional fields.

Along the way, I encountered some interesting questions.
First, what is a ``local interaction" in a field theory? Suppose that $\psi(x)$
and $\phi(x)$ are two fields, say complex functions of $x \in R^3$.
Suppose they are made to interact.
Then by ``local interaction" it would seem we mean a term appearing in the Lagrangian
containing the product:

\be
 F(\psi)(x).G(\phi)(x),\label{meanslocal}
\ee

\ni with both of the functions appearing here ``local", meaning that $F(\psi)(x)$
can be computed from knowledge of $\psi$ in a neighborhood of $x$, and ditto for $G(\phi)(x)$
and $\phi$. 
So, e.g., we might have:

\be
F(\psi)(x) = f(\psi(x),\partial \psi/\partial x_1(x), ..., \partial^2\,\psi /\partial x_1^2,.. ).
\ee

\ni where $f$ is some multivariable function.
Then a term in the Lagrangian of form:

\be
\int\,dt\,\int\,dx\,
 F(\psi)(x).G(\phi)(x),
\ee

\ni can be intrepreted as a sum of terms in which the two fields interact locally. 

By contrast, something of form:

\be
\int\,dt\,\int\,dx\,\int\,dy\,
 F(\psi)(x)\,K(x-y)\,G(\phi)(y),
\ee

\ni would presumably be intrepreted as representing nonlocal interactions between the two fields.
However, a term involving only one field of this form:

\be
\int\,dt\,\int\,dx\,\int\,dy\,
 h(\psi)(x)\,K(x-y)\,h(\psi)(y),
\ee

\ni can arise by ``integrating out" another field as we shall see,
and so need not imply any nonlocality!

Second, what is a ``particle interpretation" of a field theory? The answer would seem to
depend on the preferred theory and ideology. Immediately, in 1926, Max Born provided
such an interpretation of \Sch's wavefunction: its modulous squared is the probability
density for ``finding the particle there".\footnote{\Sch\ never agreed with this interpretation
of his wavefunction.} After the so-called ``second-quantization revolution" in the '40s,
the wavefunction had supposedly been relegated to history, but at the same time so was
particle localization, according to Steven Weinberg, \cite{WeinbergVI}. 
In Quantum Field Theory (QFT), in which complex (``c-valued") fields are replaced by Hilbert
space operators, a ``particle interpretation" seems to require a long-time appearance
of free fields that can be interpreted as, say, ``The particle is headed North with
such-and-such a momentum" superposed with the same, but North replaced by South.
But there is also the belief, incorporated by Arthur Wightman
in his axioms for QFT, \cite{SandW64},
 in local interactions.\footnote{I believe the local interactions are buried in his choice
of generating operators of form $\Lambda(f)$, where $\Lambda()$ can be a generalized function
such as Dirac's delta (or it's derivative!) and `$f$' denotes a test function. Presumably
Wightman believed the Hamiltonians or Lagrangians would be constructed from products of
these fields with localized test functions.}
Perhaps a residual belief that fields, whether complex-valued, multicomponent (as in Dirac's theory) or operator-valued, are really some peculiar description of particles interacting with each other
locally (and thus evading any atavistic appearance of Newton's action-at-a-distance at
a microscopic level) is why assumption (\ref{meanslocal}) seems a necessity of thought.

In this work I will also consider the possibility that some fields might not be sensitive
to microscopic variables, but only to macroscopic ones, e.g., to the Center-of-Mass (CoM)
of a (large) configuration. 
Thus we might entertain a multivariable \Sch\ wavefunction: $\psi(x_1,x_2,...,
x_N)$, together with a field depending only on the CoM, i.e., of form:

\be
\phi(X); \ph \hbox{where} \ph
X \= (1/N)\,\sum_{k=1}^N\, x_k.
\ee

\ni Call the former a ``Micro-field" and the latter a ``Macro-field". (I relegate discussing
whether the latter concept means a total abandonment of the ideal of local interactions,
violates Bell's proscriptions, and so on, to the last section.) 

The main thrust of the analysis
in this paper is determining whether, by incorporating but then integrating out, Macro-fields
my 2017 theory might be reproduced. Then it should become clear whether that theory
is ``nonlocal", whether it could be given a particle interpretation, and so forth.

In the following sections I recast my 2017 model to be similar to Barut's; demonstate
how to derive \Sch's equation (and my 2017) from a Lagrangian; introduce Micro- and 
Macro-fields and solve a one-dimensional model (but not the author's 2017); put forth
a general scheme and ask whether it can generate my 2017; prove a so-called ``No-Go" theorem;
and draw conclusions in a final
Discussion. 

\section{The 2017 proposal, rewritten and compared to Barut's ``Nonlinear, Nonlocal"
electrodynamics \label{2017section}}

The author's new terms took the following form. Suppose some macroscopic object
is describable by co{\"o}rdinates: $x_1,x_2,...,x_N$, with each $x_k \in R^d$. In this
section I will set $d = 1$ to make the exposition simpler to read, and examine
the options with $d = 3$ later. Let $X$ be the CoM as defined in the last section.
Then the new terms, which might be dubbed ``WaveFunction Energy" (WFE), are given by: 

\be
\hbox{WFE} \= w\,N^2\,S_N,\label{WFEdef}
\ee

\ni where `$w$' is a (presently unknown) positive parameter and
 $S_N$, the squared dispersion of $X$, is given by:

\be
S_N \= <\psi\,|\,X^2\,|\,\psi> - \left[\,<\psi\,|\,X\,|\,\psi>\,\right]^2.\label{Deq}
\ee

Note the factor of $N^2$ in (\ref{WFEdef}) which will imply that, 
in the presence of a ``cat", 
the WFE is enormous, even if the prefactor
is very small (see \cite{WickI}, \cite{Wick_Barut} for more about ``cats" and
why blocking them resolves the Measurement Problem).

There is another way to write down the same quantity. 
We ``double" the space variables; i.e.,
take another copy of $(x_1,x_2.,...,)$, call them $(y_1,y_2,...)$ and a doubled-arguments
wavefunction:

\be
\psi^{(2)} \= \psi(x_1,x_2,...)\,\psi(y_1,y_2,...),
\ee

I claim that in terms of this wavefunction we can write:

\bar
\no \hbox{WFE} &\=& (1/2)\,w\,N^2\,<\psi^{(2)}\,|\,\left[\,\left(\,\frac{1}{N}\,\right)\,\skN\,
\left\{\,x_k - y_k\,\right\}\,\right]^2\,|\,\psi^{(2)}>\\
\no &\=& (1/2)\,w\,N^2\,\int\,\prod\,dx_k\,|\psi|^2(x_1,...)\,
\int\,\prod\,dy_k\,|\psi|^2(y_1,...)\,|\,X - Y\,|^2,\\
&&
\ear

\ni where I have written similarly for the y-variables:

\be
Y \= \left(\,\frac{1}{N}\,\right)\,\skN\,y_k.
\ee

The identical formula can also be written as follows (which will be useful later on).
Let $h(x)$, resp. $h(y)$, be the (identical) marginal densities of $X$ and $Y$:

\bar
\no h(x) &\=& \int\,\prod\,dx_k\,\delta\left[\,1/N \sum\,x_k = x\,\right]\, |\psi|^2(x_1,...);\\
\no h(y) &\=& \int\,\prod\,dy_k\,\delta\left[\,1/N \sum\,y_k = y\,\right]\,|\psi|^2(y_1,...);\\
&&\label{hdef}
\ear

\ni where $\delta[\cdot]$ is Dirac's $\delta$-function. 
Then we have:

\be
\hbox{WFE} \= (1/2)\,w\,N^2\,\int\,dx\,\int\,dy\,h(x)\,h(y)\,\left[\,x - y\,\right]^2.\label{WFEeq}
\ee

After commenting on the work of A. O. Barut in atomic physics, who arrived
at a similar formula by integrating out the Maxwell component of the consistent Dirac-Maxwell
equations, \cite{Barut92}, I was motivated to explore cases differing somewhat from the one I
introduced in 2017. Thus, consider a more general case for $WFE$:

\be
\hbox{WFE} \= (1/2)\,w\,N^2\,\int\,dx\,\int\,dy\,h(x)\,h(y)\,K\left(\,x - y\,\right).
\ee

\ni for some function $K(\cdot)$ of one variable. The above formulas assume

\be
K(\,x - y\,) \= |x-y|^2.
\ee

\ni But notice that, in (\ref{WFEeq}), the cat-blocking potential is primarily due to the 
factor of $N^2$. So perhaps other cases can be interesting.

\section{The Lagrangian formulation of \Sch's equation\label{LagrangeSchsection}}

Here I derive \Sch's equation of 1926 from
a variational principle and a Lagrangian density. I postulate:

\be
\cL \= \hstm\,\skN\,\left|\,\frac{\partial \psi}{\partial x_k}\,\right|^2 \+
V(x_1,x_2,...,x_N)\,|\psi|^2 \+ (1/2)\,i\,\hbar\,\left\{\,\frac{\partial \psi^*}{\partial t}\,\psi 
- \psi^*\,\frac{\partial \psi}{\partial t}\,\right\}.\label{SchLag}
\ee

Perhaps this formulation is a bit surprising, particularly the terms with first time-derivatives,
but also that the potential $V$ seems to have the wrong sign.\footnote{This is
a case where the Lagrangian is not of form: $T - V$; but there are others,
e.g., electromagnetism plus a charged particle. See the Wikipedia page on
``Lagrangian mechanics".} Next we consider the action:

\be
L(\psi) \= \dint\,\cL(x_1,x_2,...,x_N,t),
\ee

\ni where `$A$' denotes a region in $R^N$, and a variation of the wavefunction:

\be
\psi'(x_1,...) \= \psi(x_1,...) \+ \delta\psi(x_1,...).
\ee

\ni Here $\delta\psi$ denotes any function satisfying:

\be
 d\psi \= 0\ph \hbox{on the boundary of $A$ and at} \ph t = 0 \ph \hbox{or}\ph t = T.
\ee

The equations of motion will follows from the requirement that `$L$' be
unchanged to first order in the variation, for any allowed variation.
To first order in $\delta\psi$ we find:

\bar
\no \delta L &\=& L' - L\\
\no &\=& L(\psi + \delta\,\psi) - L(\psi)\\
\no &\=&
\dint\,\left\{\,\hstm\,\skN\,\left[\,
\frac{\partial \psi^*}{\partial x_k}\,
\frac{\partial \delta\psi}{\partial x_k}\, + 
\frac{\partial \delta\psi^*}{\partial x_k}\,
\frac{\partial \psi}{\partial x_k}\,\right] \+ \right. \\
\no && \left. \left[\,\delta\psi^*\,\psi + \psi^*\,\delta\psi\,\right]\,V \+ \right.\\
\no && \left. (1/2)\,i\hbar\,\left[\,
\frac{\partial \delta\psi^*}{\partial t}\,\psi 
+ \frac{\partial \psi^*}{\partial t}\,\delta\psi  
- \psi^*\, \frac{\partial \delta\psi^*}{\partial t}
- \delta\psi^*\,\frac{\partial \psi}{\partial t}\,\right] \,\right\}.\\
&&
\ear

Next, we integrate by parts on every term containing a derivative of the variation $\delta\psi$,
obtaining: 
 
\bar
\no \delta L &\=& 
\dint\,\left\{\,\hstm\,\skN\,\left[\,
 - \frac{\partial^2 \psi^*}{\partial x_k^2}\,\delta\psi
 - \frac{\partial^2 \psi}{\partial x_k^2}\,\delta\psi^*\,\right]\right. \+\\
\no && \left. \left[\,\delta\psi^*\,\psi + \psi^*\,\delta\psi\,\right]\,V \+ \right.\\
\no && \left. (1/2)\,i\hbar\,\left[\,
- \frac{\partial \delta\psi^*}{\partial t}\,\psi 
+ \frac{\partial \psi^*}{\partial t}\,\delta\psi  
+ \psi^*\, \frac{\partial \delta\psi^*}{\partial t}
- \delta\psi^*\,\frac{\partial \psi}{\partial t}\,\right] \,\right\}.\\
&&
\ear

\ni which can be written more compactly as:

\bar
\no \delta L &\=& 
\dint\,2\,\hbox{Re}\,\left\{\,\left[\, - \hstm\,\skN\,
\frac{\partial^2 \psi}{\partial x_k^2}  
\+ \psi\,V 
\- i\hbar\,
\frac{\partial \psi}{\partial t}\,\right] 
\,\delta\psi^*\,\right\}.\\
&&
\ear

Setting $\delta L = 0$, because $\delta\psi$ was arbitrary given the boundary conditions,
the integrand must be identically zero, yielding \Sch's equation:

\be
 i\hbar\,
\frac{\partial \psi}{\partial t} \=
 - \hstm\,\skN\, 
\frac{\partial^2 \psi}{\partial x_k^2}  
\+ \psi\,V.  
\ee

My 2017 dynamical equation can also be derived in this fashion, just by adding WFE
(thought of as potential energy) containing
the expression in (\ref{Deq}) to the Lagrangian. 
Of course, it is a peculiar ``potential energy" quartic in
$\psi$, yielding a dynamical equation cubic in $\psi$.

\section{The Lagrangian set-up with Micro-fields and Macro-fields I: 
A solved one-dimensional model (but not the author's 2017) \label{Lagrangesection}}

\def\phip{\phi_{+}}
\def\phim{\phi_{-}}
\def\intminfx{\int_{-\infty}^x}
\def\intminfpinf{\int_{-\infty}^{+\infty}}
\def\intxpinf{\int_x^{\infty}}

Again remaining in one spatial dimension ($d = 1$) I introduce two additional,
fields called $\phim$ and $\phip$, which are real-valued functions of $(X,t)$,
that is, on $R^2$. I will develop the method for several choices of Lagrangian.
The simplest is the following. Let

\be
\beta \= \sqrt{(1/2)\,w\,N^2}.
\ee

Then define:

\bar
\no L(\psi,\phim,\phip) &\=& \din\,\cL_{\hbox{Sch.}}(x_1,x_2,...,x_N,t) \+\\ 
\no && \beta\,\din\,|\psi|^2(x_1,..,;t)\,
\left[\,\phim(X,t) + \phip(X,t)\,\right] \-\\
\no && (1/2)\,\int\,dt\,\int dX\,\left[\,\left(\,\frac{\partial \phim}{\partial X}\,\right)^2 
+ \left(\,\frac{\partial \phip}{\partial X}\,\right)^2\,\right].\\
&&\label{Leq}
\ear

I will impose in addition the boundary conditions:

\bar
\no \phim(X) \rightarrow 0 \ph \hbox{as}\ph X \rightarrow -\infty;\\
\no \phim(X) \ph \hbox{remains bounded as} \ph X \rightarrow +\infty;\\
\no \phip(X) \rightarrow 0 \ph \hbox{as} \ph X \rightarrow +\infty;\\
\no \phip(X) \ph \hbox{remains bounded as} \ph X \rightarrow -\infty.\\
&&\label{bcs}
\ear

All three fields are to be varied independently. For the $\phi$-fields,
this yields:

\def\hath{\hat{h}}
\be
\frac{\partial^2 \phi}{\partial X^2} \= - \beta\,h(X) = \hath(X),\label{pdeone}
\ee

\ni where $h(x)$ is given in (\ref{hdef}). 
The general solution of this PDE (called the ``one-dimensional Poisson equation")
can be obtained from the Green's function\footnote{Which can be found in the middle of
a long list published on the page entitled ``Green's function" on Wikipedia, on 7/14/2025
at 1400 hours PST, 
with the printed caution that some of the formulas given are wrong. Let the reader beware.}

\be
G(x - y) \= (x - y)\,\Theta(x - y)\, \+ x\,\alpha(y) \+ \beta(y).
\ee

\ni ($\Theta(x)$ is the ``Heaviside step function": one if $x \geq 0$ and zero otherwise.)
The meaning is that our desired solution can be obtained from the recipe:

\be
\phi(x) \= \intminfx\,dy\,(x - y)\,\hath(y) \+ x\,\intminfpinf\,dy \,\alpha(y)\,\hath(y) \+
\intminfpinf\,dy\,\beta(y).
\ee

\ni where $\alpha(y)$ and $\beta(y)$ are arbitrary functions. As one instance,
let $\alpha(y) = -1$ and $\beta(y) = y$ for all $y$, yielding

\be
\phip(x) \= \intxpinf\,dy\,(y - x)\,\hath(y), 
\ee

\ni which satisfies the second set of boundary conditions in (\ref{bcs}). Define similarily

\be
\phim(x) \= \intminfx\,dy\,(x - y)\,\hath(y), 
\ee

\ni which satisfies the first set of required boundary conditions. Moreover, assuming some
regularity conditions, these are the only solutions of (\ref{pdeone}) 
satisfying these conditions.

Note that

\be
\phim(x) + \phim(x) \= \intminfpinf\,dy\,|x - y|\,\hath(y).\label{wrongthing} 
\ee

\ni Plugging into (\ref{Leq}), we obtain (\ref{WFEeq}), but with 

\be
K(x-y) \= |x-y|,
\ee

\ni which is not the case I considered in 2017. 
It does not have the ``right" scaling, and does not lend itself to a physical interpretation
as the dispersion of the CoM  (see \cite{WickI}).

\section{The Lagrangian set-up with Micro-fields and Macro-fields II: 
 A general construction for $d = 1$ \label{Lagrangesetup}}

\def\Cinfinf{C_{\pm \infty}^{\infty}}

As the Lagrangian in the previous section did not yield the ``right" model, let us
try to generalize it by writing:

\def\dtdX{\int\,dt\,\int\,dX}

\bar
\no L(\psi,\phi) &\=& L_{\hbox{Sch.}} \+\\
\no && \dtdX\,h(X)\,\phi(X) \- (1/2)\,\dtdX\,
\left[\,\Omega\phi\,\right]^2,\\
&&
\ear

\ni where `$\Omega$' stands for a linear operator on fields, to be chosen hopefully to fulfill
 our desire. Let a superscript `$t$' on an operator mean transpose, i.e., one satisfying:

\be
\int\,dX\,\left[\,\Omega^t\,g\right]\,f \= \int\,dX\,g\,\Omega\,f,
\ee

\ni for all functions $f(X)$ and $g(X)$ satisfying suitable smoothness and boundary
conditions. 
Varying $\phi$ then yields:

\be
\Omega^t\,\Omega\,\phi(X) \= h(X).\label{cpeq}
\ee

What would produce the desired replacement for the $|X - Y|$ in (\ref{wrongthing}), namely
$(X - Y)^2$? Clearly, we would like:

\be
\Omega^t\,\Omega\,\phi(X) \= \phi'''(x).\label{desiredeq}
\ee

If such an operator exists, then (\ref{pdeone}) in the last section
is replaced by:

\be
\frac{\partial^3 \phi}{\partial X^3} \= \hath(X),
\ee

\ni and the Green's function for this equation has the form:

\be
G(x - y) \= (1/2)\,(x - y)^2\,\Theta(x - y)\, \+ x^2\,\alpha(y) \+ x\,\beta(y) + \gamma(y).
\ee

\ni Proceeding as in that chapter, with suitable boundary conditions we can have the solution:

\be
\phip(x) \= \intxpinf\,dy\,(y - x)^2\,\hath(y), 
\ee

\ni and continue as before, constructing now the 2017 term in the Hamiltonian.

So it all comes down to hoping to find a linear operator interpretable as
``the 3/2 power of the derivative operator":

\be
\Omega \= \left(\,\frac{d}{dx}\,\right)^{3/2}
\ee.

Such an operator is called a ``fractional derivative" (FD) and there have been many suggestions
going back more than a century as to its construction.\footnote{The Wikipedia page
entitled ``fractional calculus" lists around 20. 
Note: there exist papers in the math-physics literature on the topic ``Fractional \Sch's
Equation" but there is no connection to the matter discussed here.}
They are all integro-differential operators.
Here are several, written in what seems to be established notation (`$D$' means $d/dX$):

\begin{description}

\item Left-infinite Riemann-Liouville FD:

 \be
^{RL}D_{-\infty}^{3/2}f(X) \= \frac{1}{\Gamma(1/2)}\,\frac{d^2}{dX^2}\,\int_{-\infty}^X
\,(X - Y)^{-1/2}\,f(Y)\,dY.
\ee

\item Left-infinite Caputo FD: 

 \be
^{C}D_{-\infty}^{3/2}f(X) \= \frac{1}{\Gamma(1/2)}\,\int_{-\infty}^X
\,(X - Y)^{-1/2}\,f''(Y)\,dY.
\ee

\item Right-infinite Riemann-Liouville FD:

 \be
^{RL}D_{+\infty}^{3/2}f(X) \= \frac{1}{\Gamma(1/2)}\,\frac{d^2}{dX^2}\,\int_X^{+\infty}
\,(X - Y)^{-1/2}\,f(Y)\,dY.
\ee

\item Right-infinite Caputo FD: 

 \be
^{C}D_{+\infty}^{3/2}f(X) \= \frac{1}{\Gamma(1/2)}\,\int^{+\infty}_X
\,(X - Y)^{-1/2}\,f''(Y)\,dY.
\ee
\end{description}

Note that these are singular integrals and so are defined only on classes of functions
with suitable differentiability and boundary conditions. With that restriction,
there are relationships between them of interest here; e.g.,

\be
^{RL}D_{-\infty}^{3/2;t}f(X) \=
^{C}D_{+\infty}^{3/2}f(X).
\ee

Does (\ref{desiredeq}) then hold for the Riemann-Liouville FD? No. (I leave the calculation to
to the reader.) Rather than pursue the hunt among the other 20-odd suggested FDs 
for this desired result,
I present next an argument that such an operator cannot exist.\footnote{It also does not
seem that, e.g., the square of any of these claimants to be `$\sqrt{d/dx}$' gives $d/dx$,
or the square of any version of `$(d/dx)^{3/2}$' yields $d^3/dx^3$. 
A bit of false
advertising in this field.}

Suppose (\ref{desiredeq}) holds, multiply both sides by $\phi(X)$ and integrate, yielding:

\be
\int\, dX\, \left(\,\Omega\,\phi\,\right)^2 \= 
\int\, dX\, \phi\,\Omega^t\,\Omega\,\phi \= 
\int\, dX\, \phi\,\phi''' .\label{multeq}
\ee

For what class of functions do we imagine this holds? Presumably smooth functions with 
some boundary conditions.
But surely any such class will contain the class, call it $\Cinfinf$,
of functions infinitely-differentiable and which vanish together with all its derivatives at
$X = \pm \infty$.

\begin{quote}
{\bf Theorem} There cannot exist a linear operator on a function space whose domain
of definition includes $\Cinfinf$ and for which (\ref{desiredeq}) holds 
there. 

{\bf Proof}
In this domain we can integrate-by-parts (IBPs) with impunity.
IBPs thrice to obtain:

\be
\int\, dX\, \phi\,\phi''' \= - 
\int\, dX\, \phi'''\,\phi 
\ee
 
\ni Hence it is zero, hence by (\ref{multeq}) $\Omega\,\phi = 0$ identically. 
Hence $\Omega^t\,\Omega\,\phi =
\phi''' = 0$, hence $\phi$ is quadratic but vanishes at infinity, hence $\phi = 0$.
This contradicts the assumption that  $C_{\pm \infty}^{\infty} \subset \hbox{Domain}(\Omega)$.
QED.
\end{quote}

In other words, any putative ``3/2-th power of $d/dx$" satisfying (\ref{desiredeq}) 
must have the peculiar property that its domain excludes smooth functions 
vanishing at infinity which are not identically zero. 
Note that none of the candidates shown in this section could satisfy that last demand.
(I haven't checked on the other 18.)

\section{A ``No-Go" theorem in arbitrary spatial dimension\label{No-Gosection}}

Here I state a theorem (or ``theorem", since I am not going to bother about specifying function
spaces) of the infamous ``No-Go" type: meaning here, 
the claim that no theory of interacting fields
can generate the nonlinear terms in the 2017 paper by integrating out some of them.\footnote{Such
``No-Go" theorems have a bad reputation as representing instances of circular reasoning, 
as Bell pointed out
in 1966, \cite{Bell66}, which he published a year after proving one himself---the only one
I believe escapes circularity.}

So let's start with the general set-up but using variables $x,y,...$ to represent 
points in $R^d$ for any choice of spatial dimension, and letting `$dx$' or `$dy$'
represent integration over that space:

\def\dtdx{\int\,dt\,\int\,dx}

\be
 L(\psi,\phi) \= 
 \dtdx\,h(x)\,\phi(x) \- (1/2)\,\dtdx\,
\left[\,\Omega\phi\,\right]^2(x),
\ee

\ni where again $\Omega$ is some linear operator on a function space.
Varying $\phi$ gives as before:

\be
\Omega^t\,\Omega\,\phi \= h,
\ee

\ni and plugging back into $L$:

\be
L \= (1/2)\,\dtdx\,h\,\left[\,
\Omega^t\,\Omega\,\right]^{-1}\,h.
\ee

Thus the question becomes: might we have:

\be
\left[\,
\Omega^t\,\Omega\,\right]^{-1}\,h\,\=\,
\int\,dy\,||x-y||^2\,h(y),
\ee

\ni or, equivalently,

\be
h(x) \= 
\left[\,\Omega^t\,\Omega\,\right]_x\,
\int\,dy\,||x-y||^2\,h(y),
\ee

\ni where the subscript on the operator product means operating on the $x$-variable.

In other words, can the Green's function of an operator of form  
$\Omega^t\,\Omega$ be:

\be
G(x-y) \= ||x-y||^2?
\ee

Equivalently, might it be true that

\be
\left[\,\Omega^t\,\Omega\,\right]_x\,
||x-y||^2 \= \delta(x-y)\label{weirdeq}
\ee

\ni where $\delta()$ is Dirac's delta-function?

\begin{quote}
{\bf ``Theorem"}. No linear operator on a function space can possibly satisfy (\ref{weirdeq}). 
\end{quote}

{\bf Proof}. Multiply (\ref{weirdeq}) by an arbitrary function $f(y)$ in that function space,
and integrate to get:

\bar
\no f(x) &\=& \int\,dy\,f(y)\,\left[\,\Omega^t\,\Omega\,\right]_x\,
||x-y||^2;\\
\no &\=& \int\,dy\,f(y)\,\left[\,\Omega^t\,\Omega\,\right]_x\,\left[\,||x||^2\,\right] 
-2\, \left[\,\int\,dy\,f(y)\,y\,\right]^t\,\left[\,\Omega^t\,\Omega\,
\right]_x\,\left[\,x\,\right]\\ 
\no &\+& \int\,dy\,f(y)\,||y||^2\,\left[\,\Omega^t\,\Omega\,\right]_x\,1.\\
&&
\ear

\ni But this equation is absurd; no respectable space of functions would have the
property that every element is determined by a finite number of moments. QED

\section{Discussion\label{discussionsection}}

To avoid restricting the exploration to linear operators, one might entertain 
a general expression for the last term in the Lagrangian containing `normal'
derivatives. I tried:

\be
\int\,dt\,\int \,dX\,F(\phi,\phi',\phi''),
\ee

\ni but discovered that, upon varying $\phi$, the term containing $\phi'''$ dropped out.

So how should we interpret these observations? First: perhaps the scenario investigated here,
particularly the Macro-fields concept, is itself unphysical.

In his last paper, John Bell, \cite{Bell89}, 
offered an index of ``forbidden words" which he said should have 
``no place in a formulation [of physicists' theories] with any pretension to physical precision"; 
included were: {\em apparatus, microscopic, macroscopic, observable, and measurement}.
I accept this advice, but I would ask whether the Macro-fields considered fall under
his prohibition. Note that there is no scale parameter required to postulate such a beast.
Rather, my Macro-field merely 
depends on the CoM of a system which, if large, means it is insensitive
to smaller-scale perturbations, but if small, may be. (Of course, there still
remains the free parameter `$w$', which may reflect a scale.) Thus Macro-fields
are somewhat similar to the ``mean fields" that appear in certain statistical-mechanics models.

So I would give Macro-fields a pass. Perhaps the real moral of the story is that
the 2017 proposal cannot be interpreted as particle theory disguised. There really are
no `particles', meaning tiny bits of matter that interact locally with each other, 
in my version of \Schism.

\end{document}